\documentclass[submission,copyright,creativecommons]{eptcs}
\providecommand{\event}{ICLP DC 2022} 
\usepackage{breakurl}             
\usepackage{underscore}           
\usepackage{pgfplots, pgfplotstable} 
\usepackage{mathtools}
\RequirePackage{doi}

\title{Decomposition Strategies for Solving Scheduling Problems in Industrial Applications}
\author{Mohammed M. S. El-Kholany
\institute{University of Klagenfurt \\ Klagenfurt, Austria}
\institute{Cairo University \\ Cairo, Egypt}
\email{mohammed.el-kholany@aau.at}
}
\def\titlerunning{Decomposition Strategies for Solving Scheduling Problems in Industrial Applications}
\def\authorrunning{M. M. S. El-Kholany}
\begin{document}
\maketitle

\begin{abstract}
  This article presents an overview of a research study of a crucial optimization problem in the Computer Science/Operations research field: The Job-shop Scheduling Problem (JSP). The JSP is a challenging task in which a set of operations must be processed using a set of scarce machines to optimize a particular objective. The main purpose of the JSP is to determine the execution order of the processes assigned to each machine to optimize an objective. Our main interest in this study is to investigate developing decomposition strategies using logic programming to solve the JSP. We split our goal into two main phases. The first phase is to apply the decomposition approach and evaluate the proposed model by solving a set of known benchmark instances. The second phase is to apply the successful decomposition methods obtained from the first phase to solve a scheduling problem in the real-life application. In the current state, we finished the first phase and started the second one aiming to have a model that can provide a schedule of a factory for a short-time period.
\end{abstract}

\section{Introduction}

Real-life scheduling problems are very complex because of the combinatorial nature and constraints dictated by different production environments. The classical JSP has been known as a complex and combinatorial optimization problem since the 1950s and it was shown to be NP-hard in \cite{lenstra1977complexity}. Therefore, it is not easy to obtain the optimal solution. Until now, the optimal solution of some benchmark instances with medium size is unknown. In reality, the scheduling problems are much more complicated than the classical ones because many jobs should be scheduled and it may be up to $10000$ in some mechanical workshops \cite{zhang2010hybrid}. In addition, there are a lot of integrated factors that make the problem more complex. The general overview of the scheduling is to find the sequence of the tasks assigned to a scarce resource while optimizing an objective. Our study focuses on the Job-shop Scheduling Problem (JSP). The JSP is defined as a set of $n$ jobs $J_{1}, ..., J_{n}$ that have to be executed by a set of $m$ machines $M_{1}, ... M_{m}$ while finishing all of these jobs as early as possible. Each job has a series of $m$ operations with a predefined order and must visit each machine one time. Each machine can process an operation at a time. A machine cannot be interrupted once it starts processing an operation. The processing time of each operation is fixed. The naive objective of the JSP is to complete all jobs in the shortest time, which is called makespan. The main goal of solving such a problem is to find the sequence of operations assigned to each machine that completes all the jobs in the shortest possible time while satisfying the precedence constraint. Since the problem is complex and developing a model that provides an efficient schedule is challenging, many researchers have studied the JSP and proposed several techniques to solve it. The following section summarizes the previous work that tackled the JSP.

\section{Literature Review}
This section summarizes the work studied on solving scheduling problems, particularly the JSP. The scheduling problem has been addressed using exact and/or heuristics techniques. Several researchers have developed and proposed different dispatching rules to order the operations stated in front of each machine to obtain an efficient schedule in a reasonable time. Many dispatching rules have been proposed in \cite{holthaus1997efficient,kaban2012comparison,paul2015investigation} for solving different scheduling problems. These studies focused on minimizing either the total tardiness or makespan. However, the main issue with the dispatching rule is that there is no optimization process behind it. It aims to define combined rules based on the information of the problem and evaluate the obtained schedule w.r.t the performance measures. Other researchers have applied data mining techniques that study the sequence pattern of the operations assigned to each machine and develop some rules to order the operations in front of each machine \cite{shahzad2010discovering,koonce2000using}. Another group of researchers has presented exact methods such as Mixed-integer Programming models or constraint programming for solving small benchmark instances of the JSP while minimizing the tardiness \cite{ku2016mixed,meng2020mixed}. The drawback of the exact methods is that they fail to obtain a good solution in a reasonable time for large instances. This leads many researchers to follow another decomposition approach that aims to partition the problem into subparts, optimize each separately, and merge all sub-solutions to solve the main problem. Different decomposition strategies have been proposed during the last two decades. Most of the previous work decomposed the problem based on different criteria and applied heuristics/metaheuristics to solve the problem~\cite{singer2001decomposition, zhai2014decomposition,byeon1998decomposition}. This study aims to introduce new decomposition strategies to split the problem efficiently into parts (time windows) and solve them to obtain good solutions reasonably. The decomposition optimization processes are applied using a Logic Programming language called Answer Set Programming.

\section{Research Goal}
The main goal of our research is to build a model that can provide an efficient schedule for manufacturing industrial systems for a short period using Answer Set Programming (ASP). ASP is a programming methodology rooted in artificial intelligence and computational logic research \cite{lifschitz2002answer}. The main reason for selecting ASP as a solver is because it showed its effectiveness for solving different combinatorial optimization problems in several fields \cite{abseher2016shift,ricca2012team,abels2019train}. Since the scheduling problem is very complex and increasing the number of operations to a schedule makes the problem much harder. We aim to use the decomposition approach to obtain a good schedule in a reasonable time. We can split our main goal into two main steps. The first is to develop different decomposition strategies and investigate their effectiveness. Therefore, we can determine the most critical factors that significantly impact the solution quality. Secondly is to build the scheduling model using ASP, which optimizes each time window separately by applying a multi-shot ASP solving technique\cite{gebser2019multi}. 

\subsection{Accomplished work}
In order to accomplish our goal, we decided to build a scheduling model using ASP and assess its performance by solving a set of known benchmark instances. Firstly, we have built a basic model of ASP to solve the classical JSP and we found that it cannot find good solutions for the instances with more than 500 operations. So, we investigated applying the decomposition approach in which the problem is cut into sub-parts called (time windows) and the ASP optimizer solves each dynamically using the multi-shot solving technique \cite{gebser2019multi} and get the solution of the original problem. The main contribution of this approach is to find a smart way to split the problem into time windows. More specifically, determining which operations are assigned in which time window with considering the precedence constraint and obtaining a near-optimal solution in a reasonable time. We have developed/proposed different decomposition methods based on different techniques and tested them on a set of benchmark instances generated in \cite{taillard1993benchmarks,demirkol1998benchmarks}. Firstly, we applied a constrained k-Means clustering algorithm for the decomposition process. This work depends on collecting information from the problem itself and the solutions obtained by heuristics like FIFO/MTWR to define the similar operations and assign them to the same cluster where a cluster is a time window and apply multi-shot ASP for the optimization. This work has been published in PADL 2022 \cite{el2022decomposition}. Secondly, We have developed other decomposition strategies that depend on ranking the operations. Our two main decomposition ideas in that work are time/machine-based methods. The time-based approach ranks the operations based on the Earliest possible Starting Time (EST) or the Most Total Work Remaining (MTWR) and the operation with a smaller rank will be assigned to an earlier time window. For simplicity, we will describe only the EST strategy. In the EST (time-based), the EST of first operations of all jobs is $0$ because they have no predecessors. The EST of the successors is the sum of processing times for the predecessor operations belonging to the same job. The operation with smaller EST will be assigned to an earlier time window. If the EST of two or more operations is similar, the operation with a shorter processing time will be assigned to an earlier time window. However, in the EST (machine-based), the first step is to rank the operations based on EST. Secondly, we determine the bottleneck machines with the longest processing time to execute. Thirdly, we select the operation assigned to a bottleneck machine with the smallest EST; if the predecessor of that operation is not assigned to a time window, it will be assigned to that time window to satisfy the precedence constraint. This work has been accepted and will be presented in ICLP 2022. In addition, we applied different techniques to improve the solution quality. Firstly, we applied the overlapping between the time windows. This process aims to overcome the mistakes due to the decomposition. Secondly, we performed a post-processing phase (compression) after optimizing each time window. In this phase, we check the possibility of executing an operation earlier without violating the precedence constraint.

\section{Results}
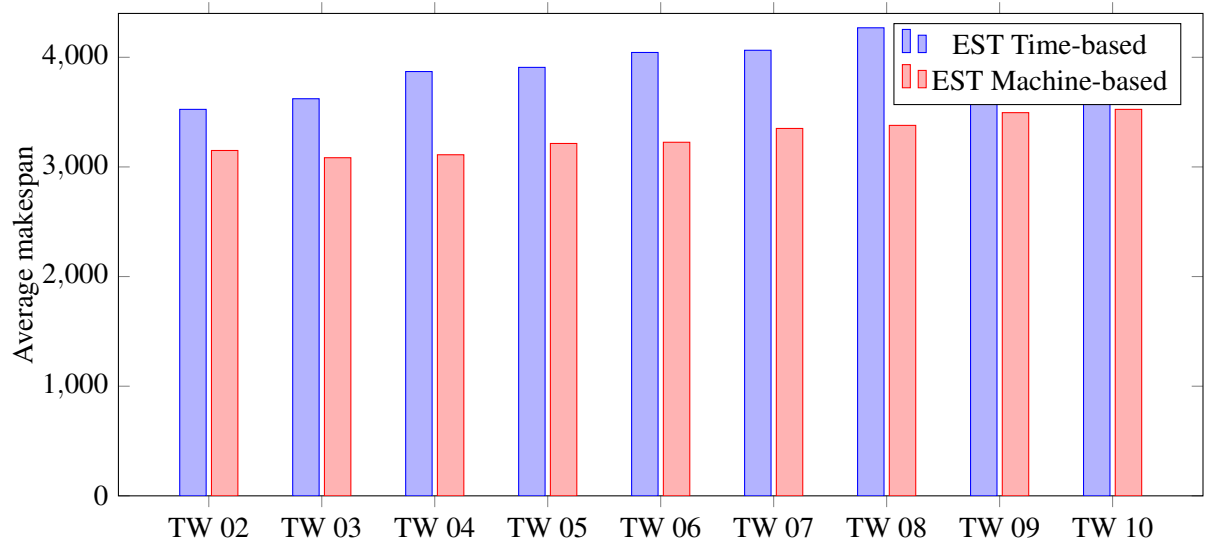
\begin{figure}
  \begin{tikzpicture}
    \begin{axis}[
                ybar,
                symbolic x coords={TW 02,TW 03,TW 04,TW 05,TW 06,TW 07,TW 08,
                TW 09,TW 10},
                width=\textwidth,
                ymin=0,
                ymax=4400,
                height=8cm,
                ylabel={Average makespan},
                xtick=data,
            ]
            \addplot table[x=interval,y=EST Time-based]{\mydata};
            \addplot table[x=interval,y=EST Machine-based]{\mydata};
            \legend{EST Time-based, EST Machine-based};
    \end{axis}
  \end{tikzpicture}
  \caption{Comparison between (Time and Machine)-based decomposition strategy}  
  \label{fig:figure01}
\end{figure}

\begin{figure}
  \begin{tikzpicture}
    \begin{axis}[
       xtick=data,
       xticklabels from table={\overlapping}{instance},      
       bar width=7mm, y=4mm,
        nodes near coords align={vertical},
        ymin = 15,
        ymax = 37,
        height = 4cm,
        width=\textwidth,
        ylabel={Gap to optimality (\%)},
      ]
    \addplot table [ybar, y=0, x expr=\coordindex,] {\overlapping};  
    \addplot table [y=10, x expr=\coordindex] {\overlapping}; 
    \addplot table [y=20, x expr=\coordindex] {\overlapping}; 
    \addplot table [y=30, x expr=\coordindex] {\overlapping}; 
    \legend{0\% , 10\%, 20\%, 30\%}
    \end{axis}
  \end{tikzpicture}  
  \caption{EST Time-based with overlapping (03 TW)} 
  \label{fig:figure02}
\end{figure}
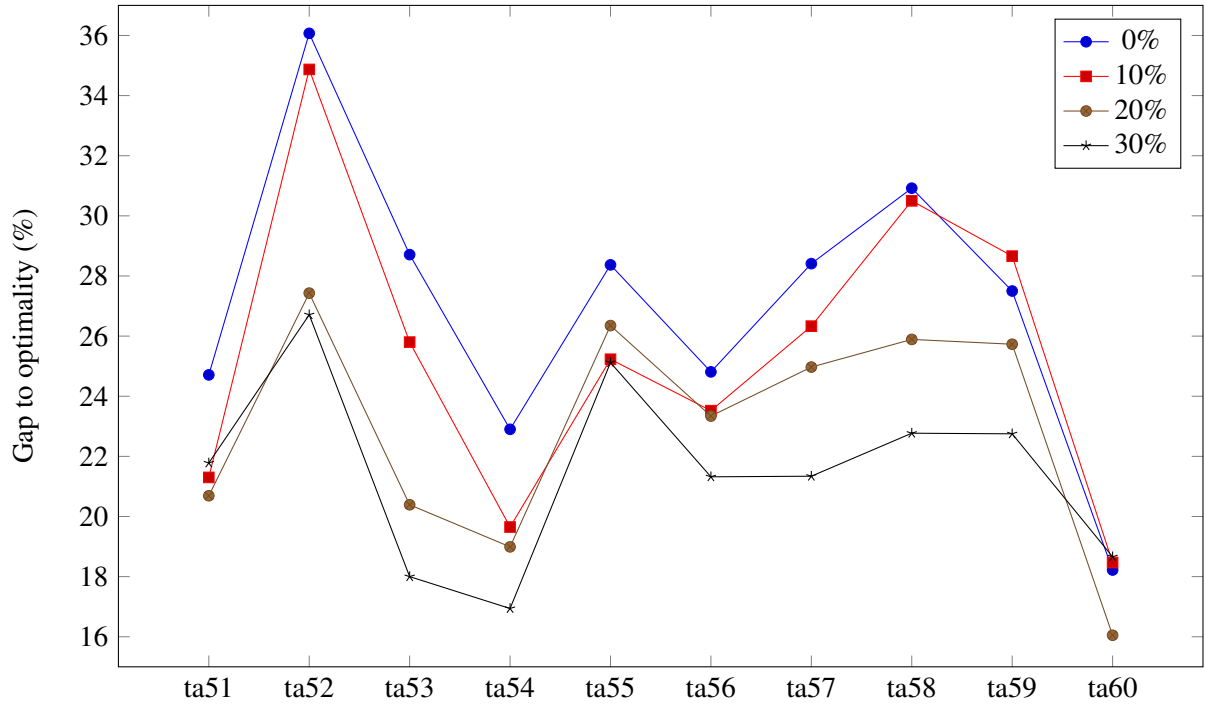

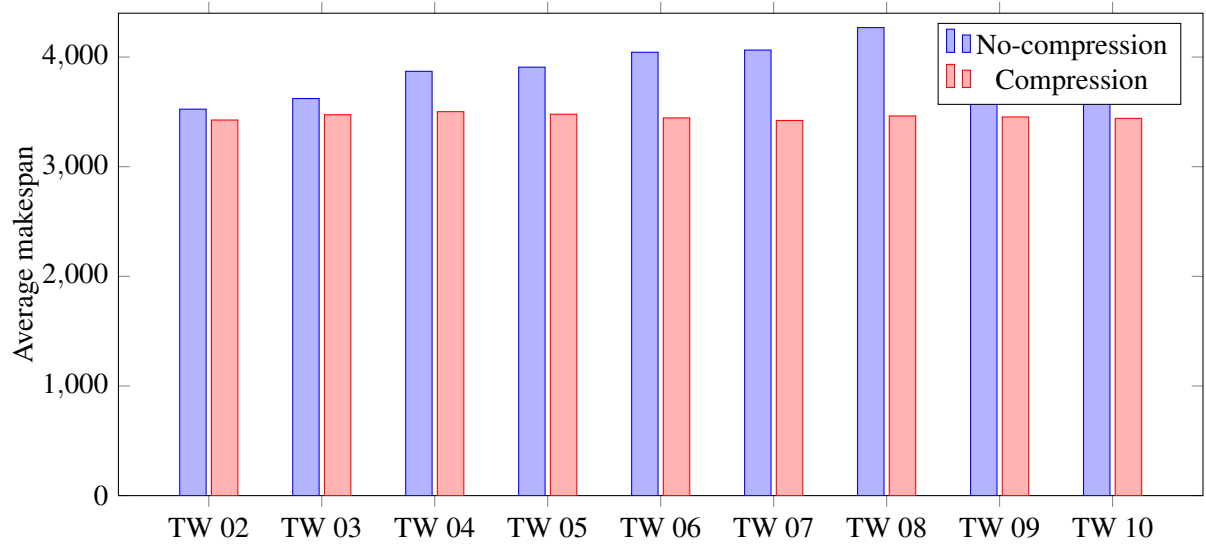
\begin{figure}
  \begin{tikzpicture}
    \begin{axis}[
                ybar,
                symbolic x coords={TW 02,TW 03,TW 04,TW 05,TW 06,TW 07,TW 08,
                TW 09,TW 10},
                width=\textwidth,
                ymin=0,
                ymax=4400,
                height=8cm,
                ylabel={Average makespan},
                xtick=data,
            ]
            \addplot table[x=interval,y=No-compression]{\compression};
            \addplot table[x=interval,y=Compression]{\compression};
            \legend{No-compression, Compression};
    \end{axis}
  \end{tikzpicture}
  \caption{Makespan EST Time-based with/without compression}
  \label{fig:figure03}
\end{figure}

This section shows a summary of the obtained results with a set of benchmark instances created in \cite{taillard1993benchmarks}. We selected the sets with the largest size which are $50\times15$, $50\times20$ and $100\times20$. We will focus only in this article on the size $50\times15$ to avoid presenting too many results of our experiments. We performed the experiments on our selected instances with different time windows ($2 - 10$) to determine the number of time windows that provide a good makespan with a timeout of $1000$ seconds. Figure ~\ref{fig:figure01} shows the average makespan of the instances set $50\times15$ with different time windows while applying EST Time-based and Machine-based decomposition methods. The results show that the Machine-based decomposition method is more effective than the Time-based method throughtout all the time windows. The best number of time windows for such a set is $3$. After determining the best number of time windows for the instances, we applied the overlapping technique with different percentages.
The overlapping percentage represents the number of operations assigned to the previous time window and will be rescheduled with the current time window. Increasing the percentage could provide better results; however, the computational time will increase. So, we run the experiments with overlapping percentages to determine the best percentage for the selected set, which provides a good result in a reasonable time. Figure ~\ref{fig:figure02} shows the makespan of each instance when the EST Time-based is applied with $03$ time windows with/without overlapping. The horizontal line shows the selected instances and the vertical shows the gap to the optimality. Each line represents a particular overlapping percentage where $0\%$ means no overlapping. We can see from the figure that increasing the overlapping percentage improves the solution quality. When we set the overlapping percentage to $40\%$ and $50\%$, the results were worse than the obtained with smaller percentages. The last step we applied to improve the solution quality is compression. The main goal of this phase is to check after optimizing the current window if it is possible to execute some operations earlier without violating the precedence constraint. Figure ~\ref{fig:figure03} shows the average makespan throughtout the time windows when we applied the compression and compares the results without compression. As shown in the figure, the compression improves the solution quality for all time windows. We can conclude from the obtained results in this section that considering the bottleneck machines (has the highest processing time to finish) during the decomposition process is beneficial. In addition, the overlapping plays a significant role in improving the solution quality because it overcomes the decomposition mistakes. Also, applying the compression phase after optimizing each time window has a positive impact. 

\pgfplotstableread[row sep=\\,col sep=&]{
    interval & EST Time-based & EST Machine-based \\
    TW 02     & 3525  & 3150   \\
    TW 03     & 3622 & 3084   \\
    TW 04     & 3870 & 3111  \\
    TW 05     & 3908 & 3214  \\
    TW 06     & 4044  & 3225  \\
    TW 07     & 4064  & 3351 \\
    TW 08     & 4269  & 3379  \\
    TW 09     & 4146  & 3495  \\
    TW 10     & 4127  & 3525  \\
    }\mydata

\pgfplotstableread[row sep=\\,col sep=&]{
instance  &  0    & 10    & 20    & 30 \\
ta51      & 24.71 & 21.30 & 20.69 & 21.78 \\
ta52      & 36.07 & 34.87 & 27.43 & 26.71 \\
ta53      & 28.71 & 25.80 & 20.39 & 18.00 \\
ta54      & 22.90 & 19.65 & 18.99 & 16.94 \\
ta55      & 28.37 & 25.23 & 26.35 & 25.12 \\
ta56      & 24.81 & 23.52 & 23.34 & 21.32 \\
ta57      & 28.41 & 26.33 & 24.97 & 21.34 \\
ta58      & 30.92 & 30.50 & 25.89 & 22.77 \\
ta59      & 27.50 & 28.66 & 25.73 & 22.75 \\
ta60      & 18.22 & 18.47 & 16.05 & 18.66 \\
}\overlapping

\pgfplotstableread[row sep=\\,col sep=&]{
    interval & No-compression & Compression \\
    TW 02     & 3525  & 3426   \\
    TW 03     & 3622 & 3474   \\
    TW 04     & 3870 & 3502  \\
    TW 05     & 3908 & 3479  \\
    TW 06     & 4044  & 3445  \\
    TW 07     & 4064  & 3422 \\
    TW 08     & 4269  & 3463  \\
    TW 09     & 4146  & 3454  \\
    TW 10     & 4127  & 3441  \\
}\compression

\section{Expected Achievement}
The classical JSP is limited and has many assumptions which make the problem unrealistic. We currently aim to apply the successful (Machine-based) decomposition strategies, overlapping and compression, to semiconductor manufacturing systems. In order to accomplish this, we are currently working on a dataset generated from a simulator called SMT2020 \cite{kopp2020smt2020}. The advantage of this dataset is that it links between the academic and industrial perspectives. Since the SMT2020 data set is quite big, it is not possible to provide a schedule with a simulator that simulates the system for $2$ years, as has been presented in \cite{kopp2020smt2020} using ASP. We aim to develop a scheduler model that could provide a schedule in the manufacturing system for a short period, like one week. The given dataset is completely different from the classical JSP, where there is a set of lots instead of jobs. Each lot belongs to a particular product, consists of steps (operations in the JSP), and is released at a particular time. Each step can be executed by different machines \textit{Flexible scheduling problem} (FSB); however, in the JSP, each operation is executed by only one machine. Since the problem is FSB, it is required to find a criterion to assign each step to a machine before starting to schedule. In addition, the optimization criterion is to minimize the total tardiness of the lots while considering other factors such as machine maintenance, cascading, and batching. Given that information, our plan is:
\begin{enumerate}
  \item Firstly, we aim to build a basic ASP model that assigns each step to a machine and then schedules the lots considering only the machine maintenance.
  \item Secondly, develop decomposition strategies \textit{(machine-based)} and apply the multi-shot solving ASP.
  \item Thirdly, we will consider the other mentioned factors.
\end{enumerate}

In order to assess our model, we will compare our obtained results with the results of colleagues (in our research group) who are working on the same dataset and applying Deep Learning and dispatching rules.

\nocite{*}
\bibliographystyle{eptcs}
\bibliography{generic}
\end{document}